\documentclass[prb]{revtex4}
\usepackage{graphicx}
\usepackage{dcolumn}
\usepackage{bm}

\begin{document}

\title{Short chains at solid surfaces: wetting transition from a density functional approach}
\author{P. Bryk}
\affiliation{Department for the Modeling of Physico-Chemical Processes,
Maria Curie-Sk{\l}odowska University, 20-031 Lublin, Poland}
\author{S. Soko{\l}owski}
\affiliation{Department for the Modeling of Physico-Chemical Processes,
Maria Curie-Sk{\l}odowska University, 20-031 Lublin, Poland}
\begin{abstract}
A microscopic density functional theory is used to investigate the adsorption of short
chains on attractive solid surfaces. We analyze the structure of the adsorbed fluid
and investigate how the wetting transition changes with the change of the chain length and
with relative strength of the fluid-solid interaction. 
End segments adsorb preferentially in the first adsorbed layer
whereas the concentration of the middle segments is enhanced in the second
layer. 
We observe that the wetting temperature rescaled by the bulk critical
temperature decreases with an increase of the chain length. For longer chains 
this temperature reaches a plateau. For the surface critical temperature
an inverse effect is observed, i.e. the surface critical temperature
increases with the chain length and then attains a plateau.
These findings may serve as a quick estimate of the wetting and surface critical
temperatures for fluids of longer chain lengths. 

\end{abstract}

\maketitle




\section{Introduction}

Alkanes are often used as standard substances in studies of wetting
and adhesion. This is  connected with the
industrial importance of these chemicals, being 
major constituents of fuels, oils and lubricants.
Consequently, the literature aiming at theoretical description of
adsorption of chain particles is very rich -- see
for the review Refs.\cite{Milchev:00, Muller:03}. Among different theoretical
approaches based on application of lattice-off models,
three groups of methods are of a particular importance. These
are computer simulation techniques, 
\cite{Milchev:00,Smith:00,Balasubramanian:95}, 
the self-consistent field theory calculations \cite{Muller:03,Koopal:96:1,Koopal:96:2}
and the approaches based on the applications of different versions of density
functional theory (DFT)
\cite{Sweatman:03:1,Woodward:91:1,Woodward:92:1,Yethiraj:95:1,Yethiraj:98:1,
Chandler:86:1,Hooper:00:1,Pablo:99:1,Frink:02:1,Kierlik:92:1,Yu:02:1}. 
 
In recent years there have been proposed several  density functional
approaches for molecular systems. One of the earliest density functional-based 
studies were carried out by Woodward  \cite{Woodward:91:1,Woodward:92:1} more
than ten years ago. Next, Yethiraj and Woodward  
\cite{Yethiraj:95:1,Yethiraj:98:1}
proposed a theory that combined a single-chain Monte 
Carlo simulation and a DFT scheme. 
Another class of density functional methods was based on on
the interaction site formalism of Chandler et al. \cite{Chandler:86:1,Hooper:00:1}.
Finally, we should mention about theories which  extends
Wertheim's thermodynamic perturbation theory 
\cite{Wertheim:87:1} to the case of nonuniform chain fluids
\cite{Kierlik:92:1,Yu:02:1}. From computational
point of view the latter theories, especially the theory of Yu and Wu 
\cite{Yu:02:1,Yu:03:1},
seem to be particularly convenient to study the
structure and thermodynamic properties of nonuniform fluids
composed of tangentially bonded chains.

Majority of  previous density-functional based studies of adsorption of
chain particles have considered models with repulsive forces only and
have been devoted to study of the packing of hard chains at hard walls
cf. eg. \cite{Kierlik:92:1,Woodward:91:1,Woodward:92:1,Yethiraj:95:1,Hooper:00:1,Yu:02:1,
Bryk:04}.
More recently theoretical effort has been
shifted towards studies of systems with attractive interparticle
 and particle-wall forces 
\cite{Patra:00,Patra:03,Muller:00,Muller:01,Muller:03:1,Zhang:04}.
Particularly worth to notice is the recent publication of  M\"uller et al.
\cite{Muller:00,Muller:03:1}, who have reported the
results of Monte Carlo, density functional and self-consistent field theory 
investigations of  surface and
interfacial properties of a molecular fluid composed a short linear chains
(built of 10 segments). Their approach
has been based on a model, according to which the molecules are
treated as spherical sites connected by springs and with site interacting
with other sites and with the surface via Lennard-Jones type 
potentials. Wertheim's thermodynamic
perturbation theory has been applied to model connectivity of the segments.
They have found \cite{Muller:03:1} that for the theory to be accurate, it is important
to decompose the free energy into repulsive and attractive contributions
with different approximations (e.g. different weighted densities) for the
two parts. The developed approach has been able to reproduce
the results of computer simulations with a good accuracy. However,
it also requires much of computational effort. Consequently,
extensive studies of surface phase transitions within the framework
of the theory of M\"uller et al. \cite{Muller:03:1} would be rather time-consuming.
In our opinion an attractive alternative for performing such studies provides 
theory of Yu and Wu \cite{Yu:02:1,Yu:03:1}. 

In this work we apply the  density functional theory of Yu and Wu
\cite{Yu:02:1,Yu:03:1} to study adsorption of molecules built of  
freely jointed tangent Lennard-Jones segments on a solid surface. The main
purpose of our work is to investigate the wetting transition
for short chains  built of from 2 up to 10 identical segments. 
A mean-field approximation is applied to calculate 
the contribution to the free energy functional arising form
segment-segment interaction. Therefore we do not expect 
our theoretical predictions to remain in a quantitative
agreement with computer simulations. However, we can expect that
a general picture of the wetting phenomena in such systems should
be correctly predicted at a qualitative level.

The evaluations of surface phase diagrams requires the knowledge
of the bulk phase diagrams for the fluids in question. Numerous calculations
of the bulk thermodynamic properties for models similar to that applied here 
have been reported in the literature, see e.g.
\cite{Muller:03:1,Villegas:97,Dowell:00,Lee:01,Pamies:02,Wang:01}
However, despite this similarity, there also exist several differences,
 which lead to differences in the course of the liquid-vapor coexistence envelope.
Therefore our surface calculations are preceded by the 
study aiming at the evaluation of the bulk phase equilibria.

\section{Theory}

\label{sec:theory} We consider a fluid of chain particles in contact with
an impenetrable solid surface. 
Each individual particle of the fluid is composed of $M$
spherical segments. 
Although the theory can be written down for a quite 
general case in which the chain may
consist of segments of different sizes, through this work we restrict
ourselves to the case of the segments of
identical diameters $\sigma$.
The segments are tangentially jointed by a bonding potential
acting between the adjacent segments of the same
chain.
The total bonding potential $V_b(\mathbf{R})$ is a sum of
bonding potentials $v_b$ between the segments 
\begin{equation}\label{eq:1}
V_b(\mathbf{R})=\sum_{j=1}^{M-1}v_b(|\mathbf{r}_{j+1}-%
\mathbf{r}_j|)\;.
\end{equation}
where  $\mathbf{R}\equiv (\mathbf{r}_1,\mathbf{r}_2,\cdots ,\mathbf{r}_M)$
denotes a set of coordinates describing the segment positions. 
According to our treatment this potential satisfies the relation 
\begin{equation}\label{eq:2}
\exp [-\beta V_b(\mathbf{R})]=\prod_{j=1}^{M-1}\delta (|\mathbf{r%
}_{j+1}-\mathbf{r}_j|-\sigma )/4\pi \sigma^2\;\;.
\end{equation}
Note that the model used here differs from those applied in Refs.
\cite{Muller:03,Muller:00,Muller:03:1,Dowell:00}, where
the so-called finitely extensible nonlinear elastic potential,
responsible for the connectivity between the adjacent segments
has been employed. In our approach we assume that the bonding potential, $v_b$, 
is of infinitely short range.

The fluid is in contact with a solid surface and each segment $j$ interacts 
with the surface via a Lennard-Jones (9-3) potential
\begin{equation}\label{eq:3}
v_j(z_j)=\varepsilon_{gs}\left[(z_0/z_j)^9-(z_0/z_j)^3\right],
\end{equation}
where $z_j$ is the distance between the segment $j$ and the surface. The energy
parameter, 
$\varepsilon_{gs}$, 
and distance parameter, $z_0$, are the same for all the segments.

Similar to previous studies \cite{Yu:02:1,Bryk:04} we assume that the segments have a hard core
forbidding any configuration that would lead to an overlap between two segments.
An additional pair-wise attractive interaction
between the segments is imposed. This attractive potential between
two segments separated by the distance $r$ is 
\begin{equation}\label{eq:4}
u(r)=\left\{
\begin{array}{ll}
0 & {\ \ \rm for \ \ } r\le \sigma, \\
4\varepsilon\left[(\sigma/r)^{12}-(\sigma/r)^6\right] & {\ \ \rm for \ \ } \sigma<r\le r_{cut}, \\
0 & {\ \ \rm for \ \ } r>r_{cut},
\end{array}
\right..
\end{equation}
In the above $r_{cut}$ is the cut-off distance, and
we have set its value to $3\sigma$. The energy parameter $\varepsilon$ is
independent of the interacting segment index.
The form of the potential (\ref{eq:4}) excludes the interaction between the adjacent segments
within the same chain.

Within the framework of the theory of Yu and Wu \cite{Yu:02:1} the grand
potential of the system $\Omega $ is defined as a functional of the local
density of the fluid, $\rho(\mathbf{R})$,
\begin{equation}\label{eq:5}
\Omega [\rho (\mathbf{R})]=F_{id}[\rho(
\mathbf{R})]+ 
F_{ex}[\rho (\mathbf{R})]+\int \!d%
\mathbf{R}\rho(\mathbf{R})(V_{ext}(\mathbf{R})-\mu )+F_{att}[\rho (\mathbf{R})]\;,
\end{equation}
where $\mu $ is the chemical potential of the fluid, $F_{ex}$ and $F_{id}$ are the excess
free energy of hard-sphere chains and the ideal free energy,
respectively, and $F_{att}$ is the free energy due to attractive forces between particles.
$V_{ext}(\mathbf{R})$ is the external potential for a chain molecule
assumed to be a sum of the external potential energies imposed on individual segments 
\begin{equation}\label{eq:6}
V_{ext}(\mathbf{R})=\sum_{j=1}^{M}v_j(\mathbf{r}%
_j)\;.
\end{equation}

The configurational ideal part of the free energy functional, $F_{id}$ is
known exactly 
\begin{equation}\label{eq:7}
\beta F_{id}[\rho(\mathbf{R})%
]=\left\{ \beta \int \!\!d\mathbf{R}\rho (\mathbf{R}) V_b(\mathbf{R})\right. 
\left. +\int \!\!d\mathbf{R}\rho (\mathbf{R})\ln (\rho (\mathbf{R%
}))-1]\right\} \;.
\end{equation}
Following Yu and Wu \cite{Yu:02:1} we assume that $F_{ex}$
is a functional of  average segment density defined as 
\begin{equation}  \label{eq:8}
\rho _s(\mathbf{r})=\sum_{j=1}^{M}\rho _{s,j}(\mathbf{r}%
)=\sum_{j=1}^{M}\int \!\!d\mathbf{R}\delta (\mathbf{r}-\mathbf{r}%
_j)\rho(\mathbf{R})\;,
\end{equation}
where $\rho _{s,j}(\mathbf{r})$ is the local density of the segment $j$
of the chain. 

The free energy $F_{ex}$ is a volume integral over free energy density,
$F_{ex}=\int \Phi (\mathbf{r})d\mathbf{r}$, 
and the free energy density can be split into the hard-sphere
contribution, $\Phi_{(HS)}$, resulting from the hard-sphere repulsion
between  segments and the contribution due to the chain connectivity, 
$\Phi _{(P)}$.

For the hard-sphere contribution we use the following expression, resulting
from the fundamental measure theory  \cite
{Rosenfeld:89:1,Roth:02:1,Yu:02:2} 
\begin{equation}  \label{eq:9}
\Phi _{(HS)}=-n_0\ln (1-n_3)+\frac{n_1n_2-\mathbf{n}_1\cdot \mathbf{n}_2}{%
1-n_3}+n_2^3(1-\xi ^2)^3\frac{n_3+(1-n_3)^2\ln (1-n_3)}{36\pi n_3^2(1-n_3)^2}%
\;,
\end{equation}
where $\xi (\mathbf{r})=|\mathbf{n}_2(\mathbf{r})|/n_2(\mathbf{r})$.

The contribution $\Phi _{(P)}$ is evaluated using Wertheim's first-order
perturbation theory \cite{Wertheim:87:1} 
\begin{equation}  \label{eq:10}
\Phi _{(P)}=\frac{1-M}{M}n_0\zeta\ln
[y_{(HS)}(\sigma)]\;,
\end{equation}
where $\zeta =1-\mathbf{n}_{V2}\cdot \mathbf{n}%
_{V2}/(n_2)^2$, with $y_{(HS)}$ given by the
Carnahan-Starling 
expression for the contact value of the radial distribution function of
 hard spheres is used
\begin{equation}  \label{eq:11}
y_{hs}(\sigma )=\frac 1{1-n_3}+\frac{n_2\sigma\zeta }{%
4(1-n_3)^2}+\frac{(n_2\sigma)^2\zeta }{72(1-n_3)^3}\;.
\end{equation}
with $\zeta =1-\mathbf{n}_{V2}\cdot \mathbf{n}_{V2}/(n_2)^2$. In Eqs.~[(%
\ref{eq:9})-(\ref{eq:11})] the spatial dependence of all variables has been
suppressed, for the sake of simplicity.
In the above $n_\alpha $, $\alpha =0,1,2,3,V1,V2$ denote the weighted densities defined as
the spatial convolutions of the average densities and the corresponding weight functions 
\begin{equation}  \label{eq:12}
n_\alpha (\mathbf{r})=\int \!\!d\mathbf{r}^{\prime }\rho _s(\mathbf{r}^{\prime
})w_\alpha (\mathbf{r}-\mathbf{r}^{\prime })\;,
\end{equation}
with the weight functions $w_\alpha (\mathbf{r})$ given in Ref.~\cite
{Rosenfeld:89:1}.

Finally, the mean-field attractive potential contribution
to the free-energy is given by
\begin{equation}\label{eq:13}
F_{att}=\sum_{j,j'=1,M}
\frac{1}{2}\int d{\bf r}d{\bf r'}u(|{\bf r}-{\bf r'}|)
\rho_{s,j}({\bf r})\rho_{s,j'}({\bf r'}).
\end{equation}
If all the segments are identical, Eq. (\ref{eq:13}) becomes
\begin{equation}
F_{att}=
\frac{1}{2}\int d{\bf r}d{\bf r'}u(|{\bf r}-{\bf r'}|)
\rho_{s}({\bf r})\rho_{s}({\bf r'}).
\end{equation}

At equilibrium the density profile $\rho(\mathbf{R}%
)$ satisfies the condition 
${\delta \Omega [\rho (\mathbf{R})]}/ \delta \rho (\mathbf{R})=0$.
This condition leads to the equation 
\begin{equation}  \label{eq:14}
\rho (\mathbf{R})=\exp \left\{ \beta \mu -\beta V_b(%
\mathbf{R})-\beta \sum_{j=1,M}\lambda _j(\mathbf{r}_j)\right\}\;,
\end{equation}
where $\lambda _j(\mathbf{r}_j)$ is 
\begin{equation}\label{eq:15}
\lambda _j(\mathbf{r}_j)=\frac{\delta \left[ F_{ex}+F_{att}\right]}{\delta \rho _s(%
\mathbf{r}_j)}+v_{j}(\mathbf{r}_j).
\end{equation}

The density profile equation (\ref{eq:14}) can be combined
with Eq.~(\ref{eq:8}) yielding the equation for the average segment local density
\begin{equation}\label{eq:16}
\rho _s(\mathbf{r})=\exp (\beta \mu )\int d\mathbf{R}%
\sum_{j=1}^{M}\delta (\mathbf{r}-\mathbf{r}_j)\exp \left[ -\beta
V_b(\mathbf{R})-\beta \sum_{l=1}^{M}\lambda _l(\mathbf{r}%
_l)\right] \;.
\end{equation}

If the density distribution varies only in the $z$-direction the
last equation can be rewritten as 
\begin{equation}  \label{eq:17}
\rho _s(z)=\exp(\beta \mu )\sum_{j=1}^{M}\exp [-\beta
\lambda
_j(z)]G_j(z)G_{M+1-j}(z)\;,
\end{equation}
where the propagator function $G_j(z)$ is determined from the recurrence relation 
\begin{equation}  \label{eq:18}
G_j(z)=\int dz^{\prime }\exp [-\beta \lambda _j(z^{\prime})]%
\frac{\theta (\sigma -|z-z^{\prime }|)}{2\sigma }%
G_{j-1}(z^{\prime })
\end{equation}
for $j=2,3,\dots ,M$ and with $G_1(z)\equiv 1$. 

\section{Results and Discussion}
We have studied   adsorption of
chains composed of $M=2$, 4, 6, 8 and 10 segments. Prior to surface studies, we
have evaluated the bulk phase diagrams
using the theory outlined in the preceding section with the local
density being constant and equal to the bulk density, $\rho_b$.
In a homogeneous system the vector weighted densities vanish, while the
scalar weighted densities become proportional to the bulk
density. From Eq.~(\ref{eq:12}) we obtain $%
n_{\alpha}=\xi_{\alpha} M\rho_b=
\xi_{\alpha}\rho_{s,b}$, with $\xi_3=(%
\sigma)^3\pi/6$, $\xi_2=(\sigma)^2\pi$, $%
\xi_1=\sigma/2$ and $\xi_0=1$. Insertion of the bulk
weighted densities defined above
into Eqs.~(\ref{eq:9})-(\ref{eq:11}) together
with Eq.~(\ref{eq:7}) leads to the total free energy per
unit volume, from which the pressure and the chemical potential
can be readily obtained. The bulk phase diagrams 
were obtained from the condition that 
at the coexistence the chemical potentials and
the pressures of a fluid in the liquid and vapor phases
are equal. As pointed out in Ref. \cite{Muller:03}, Wertheim thermodynamic perturbation
theory does not accurately describe the gas phase but it relatively
well accounts for the properties of a liquid phase of provided
that the system is well removed from the critical region.

In Fig. 1 we show examples of the liquid-vapor coexistence
envelopes in the reduced segment density 
$\rho_{b}^*=\rho_{s,b}\sigma^3$ -- reduced temperature
$T^*=k_BT/\varepsilon$ plane for several values of the chain length $M$. 
Although the surface calculations have been carried out for $M\le 10$,
the result for $M=16$ is also included
here. Diamonds mark the critical points
for all the chains lengths from $M=2$ up to $M=16$.
For $2 <\le M \le 10$ the critical temperatures
can be approximated by  $T^{*,cr} = 0.75764 + 0.95646 \ln(M)$.
The corresponding correlation coefficient is high and equals 0.99948.
Moreover, the critical densities can be approximated by
$\rho_{s,b}^{cr} = 0.26376 M^{-0.38252}$ with the correlation coefficient
equal -0.99955. The knowledge of bulk phase diagrams is necessary
for the studies of adsorption from gaseous bulk phase
at temperatures lower than the bulk critical temperature. 

We begin with the presentation of the results obtained for 
$\varepsilon_{gs}/\varepsilon=6$.
For a given value of $M$ we first solve the density profile equation
(\ref{eq:17}) and then evaluate the "segment adsorption isotherm"
\begin{equation}
\Gamma=\int_0^{ \infty } dz[\rho_s(z)-\rho_{s,b}]
\end{equation}
and the excess grand potentials $\Delta \Omega=\Omega - \Omega_b$, where 
$\Omega_b$ is the grand potential of an uniform fluid being in
equilibrium with the nonuniform system. The analysis of the dependence
of $\Delta \Omega$ on the chemical potential allows us for a precise
location of the surface phase transitions. 
 Note that the form of the external potential
Eq.~\ref{eq:3} implies that the resulting wetting transition is first order
\cite{Dietrich:88,Sullivan:86}.

In figure 2  we show examples of the adsorption isotherms, obtained
for $M=8$ at two temperatures. The first of them, $T^*=1.1$
(cf. Fig. 2a) is somewhat higher than the
wetting temperature. For the system in question the latter temperature 
is approximately equal to $T_w^*\approx 0.845$.
The adsorption isotherm in Fig. 2a  exhibits a 
well-pronounced prewetting jump. This jump occurs at the bulk density 
significantly lower than the segment density at bulk coexistence.
Instantaneously, the adsorption before the jump is very low. 
Indeed, it is almost zero on the figure scale.

With the temperature increase the prewetting jump becomes
smaller, see Fig. 2b where we have plotted
the isotherm evaluated  at $T^*=1.85$. This temperature  is close 
to the surface critical temperature, which for the investigated
system equals to $T^*_{sc}\approx 1.965$. We have also shown here
the metastable branches of the isotherm. The equilibrium transition
is marked by dashed line.

Figure 3 shows representative examples of the average segment
density profiles, calculated for the systems from Fig. 2.
Profiles in Fig. 3a have been
evaluated at $T^*=1.1$ for the bulk fluid densities lower (dashed lines) and
higher (solid line) than the density at the prewetting jump.
Dotted line marks the bulk segment density for the corresponding liquid
phase. The profile at the bulk density lower
than the density at the prewetting transition is extremely
low (the relevant curve in Fig. 3a has
been multiplied by 500) and extends over approximately one
layer. After the prewetting a thick film exhibiting
a well-pronounced layered structure and extending over approximately 6
monomer layers is developed. The film thickness is lower
than the chain length and the consecutive maxima of the profile are at 
a distance approximately equal to $\sigma$. Similar thick film
morphology is found in simple fluids, however in the chain fluid
considered here the first peak is relatively smaller.

Figure 3b illustrates the film growth after the prewetting
transition at $T^*=1.85$. Note that, contrary to
Fig. 3a, the profile before the transition (dashed line)
has not been magnified. This is reflected in the smaller jump
on the corresponding adsorption isotherm (cf. Fig. 2b).
Directly after this transition, the surface film extends over approximately
three layers. At the bulk density close to the bulk coexistence
the film thickness grows rapidly, and at the highest density displayed in
Fig. 3b spans approximately 18 monomer layers. The
layered structure of the film is less pronounced than at $T^*=1.1$.
Indeed, only three consecutive maxima on
the profiles are observed. At distances from the surface larger than
$4\sigma$ the thick film exhibits a plateau within which  
the adsorbed fluid segment density is close to the bulk 
liquid density at the bulk gas-liquid coexistence. 

Figure 4 gives a further insight into the film growth.
We find that at both temperatures, $T^*=1.1$  (Fig. 4a) 
and $T*=1.85$  (Fig. 4b) the adsorption 
divergence upon approaching coexistence is a power law
i.e. $\Gamma^*\propto(\rho^*_{b,c}-\rho^*_{b})^{-1/3}$,
where $\rho_{b,c}$ is the gas segment density at bulk coexistence.
This behavior is clearly visible on the log-log plot of adsorption
where the straight line slope equals -1/3.
The same behavior has been observed for all the systems under study
and is characteristic for the complete wetting regime
in systems with long-range fluid-fluid and/or fluids-solid potentials
\cite{Sullivan:86}.

Within the present approach it is possible to track down
the density profiles of particular segments of a chain molecule.
The segment local densities provide an additional insight
into the structure of adsorbed fluid. In Fig. 5 
we show representative examples of the end ($j=1$ or, equivalently $j=8$)
and middle ($j=4$ or, equivalently $j=5$)
segment local densities obtained for $M=8$ at $T^*=1.1$
(part a) and at $T^*=1.85$ (part b).
These plots correspond to the systems from Fig. 2.
At lower temperature (cf. Fig. 5a) the shape of the end
and middle segment profiles is almost identical before the prewetting
transition due to the fact that the density is very low (the profiles
were multiplied by 500). After the prewetting transition we find that 
the consecutive maxima of both profiles are at almost the same distances
from the surface but the first local density maximum of the 
end segments is higher than that of the middle segments.
End segments seem to be more ``pinned'' to the surface.
These findings are in accordance with results of Monte Carlo simulations
and self consistent field theory \cite{Muller:03:1}.
We also note that the structure of the tails of the profiles
is similar to that observed previously in the case of a fluid-fluid
interface of a mixture of polymers built of tangentially jointed hard
spheres \cite{Bryk:04}.

Examples of the surface phase diagrams evaluated for
dimers (part a) and 8-mers (part b) are plotted in Fig. 6.
Similar to the case of simple fluids the prewetting line (solid line)
joins the binodal (dashed line) tangentially in a wetting point
(open triangle) and ends at the surface critical point (open circle).
The reduced wetting temperature $T^*_w$ and the surface critical
temperature $T^*_{sc}$ increases with an increase of the chain length.
However, since the bulk critical temperature increases as well, it is
convenient to re-scale (divide) both $T^*_w$ and $T^*_{sc}$ by
$T^*_c$. In Fig. 7 where we display the rescaled wetting
 temperature, $t^*_w\equiv T_w^*/T_c^*$, and the rescaled
surface critical temperature $t^*_{sc}\equiv T_{sc}^*/T^*_c$,
for $M=2$, 4, 6, 8 and 10. All the calculations were carried out for
$\varepsilon_{gs}/\varepsilon = 6$. For a better visualization
of the occurring changes we also plot here the bulk critical temperatures
(marked on the right-hand side $y$-axis). The increase of the chain length leads to a
decrease of  $t^*_w\equiv T_w^*/T_c^*$ and to an increase of $t^*_{sc} \equiv T^*_{sc}/T^*_c$. 
For $M\ge 8$ the both $t^*_w$ and $t_{sc}^*$ attain plateau. This information
may be useful in finding a quick estimate of the wetting or surface critical temperatures
for fluids composed of longer chain lengths.

Finally, we address briefly the issue of the surface wettability changes 
with change of the strength of the 
segment-surface potential, $\varepsilon_{gs}$.
Figure 8 presents the plot of the dependence of $T_w^*$ and $T_{sc}^*$ 
on $\varepsilon_{gs}^*=\varepsilon_{gs}/\varepsilon$ for the chains of the length $M=4$. 
When $\varepsilon^*_{gs}$ decreases both characteristic surface
temperatures increase and the gap between them becomes smaller, as expected.
For $\varepsilon^*_{gs}=2$ the surface 
critical temperature $T_{sc}^*\approx 2.01$ becomes only
marginally higher than the wetting temperature, $T_w^*\approx 2.03$ 
and both these temperatures
become only slightly lower than the corresponding bulk critical temperature.
Extrapolation of the obtained results suggest that the line of
the wetting points (triangles) would cross $T_w^*=T_c^*$
for $\varepsilon_{gs}^*\approx 1.8$. However
such decrease of $\varepsilon_{gs}^*$ would lead to a 
crossover from a complete wetting regime to a critical adsorption regime.
Since in our studies we employ a simple mean-field functional
we have not performed the relevant calculations here and leave this issue
for a more elaborated approach.
On the other hand, an increase of $\varepsilon_{gs}$ would lead
to a decrease of both $T^*_{sc}$ and $T_w^*$ temperatures. One
can also expect the occurrence of layering-type transitions at
high values of $\varepsilon_{gs}^*$ and at low temperatures.
We have not explored the layering transitions in the present work.
The relevant results will be presented in a future.

\section{Summary}
\label{sec:summary} We have carried out studies of wetting 
behavior of fluids composed of chains built of freely jointed tangent Lennard-Jones monomers.
The difference in the chain lengths leads to the difference in the wetting behavior.
We have found that the wetting temperatures rescaled by the bulk critical
temperature decrease with an increase of the chain length. For higher values of 
$M$ the observed decrease becomes smaller and the curve of the chain length dependence
of the rescaled wetting temperature, $t^*_{w}(M)$ attains a plateau for $M\ge 9$. 
In the case of the chain length dependence of the rescaled surface critical
temperatures, $t^*_{sc}(M)$ an inverse effect has been observed, i.e. this temperature
increases with the  chain length and then attains a plateau for $M\ge 9$.
We have also found that the thick film diverges according to the power law 
$\Gamma \propto (\rho_{bc}-\rho_{b})^{-1/3})$ in accordance with the
general predictions for the fluids interacting with long range forces \cite{Sullivan:86}.
The theory allows for a detailed investigation of the adsorbed fluid structure. 
Our analysis has revealed that the end segments adsorb preferentially in the
first adsorbed layer which is in qualitative agreement with the results
of Monte Carlo simulations and self consistent field theory \cite{Muller:03:1}.

It would be of interest to check our predictions against
computer simulations. However such task would require rather long-lasting calculations,
and we postpone such studies for a future.
Our study cannot be considered as complete because we have not discussed the layering 
transitions transition. However, ascertaining the global surface phase diagrams 
for a system of freely jointed Lennard-Jones chains poses many difficulties and we postpone
systematic investigations of this issue for further works. 

\begin{acknowledgments}
P.B. acknowledges financial support from the Rector of the Maria 
Curie-Sk\l odowska University, Lublin. S.S. acknowledges EU for a
partially funding this work as a TOK contract 509249. 
\end{acknowledgments}

\vspace*{0.5cm}
\centerline{\bf Figure Captions}

\textbf{Fig. 1.} Bulk phase diagrams (binodals) of tangentially
jointed Lennard-Jones chains, characterized by different number of beads
$M=2,4,8,16$ evaluated in the reduced bulk segment density $\rho^*_b$
-- reduced temperature $T^*$ representation. 
Open diamonds indicate the locations of the critical points
for chains with $2\leq M\leq 16$. 
The reduced units are defined as: $\rho^*_b=\rho_{s,b}\sigma^3$
and $T^*=k_BT/\varepsilon$.

\textbf{Fig. 2.} Segment adsorption isotherms of tangentally
jointed Lennard-Jones 8-mers adsorbed on a solid planar surface
with $\varepsilon_{gs}^*=6$ plotted as a function
of a relative undersaturation $\rho^*_b-\rho^*_{b,c}$
evaluated at two reduced temperatures: $T^*=1.1$ (part a), and $T^*=1.85$
(part b). Dashed line in part b marks the equilibrium prewetting
transition.

\textbf{Fig. 3.} (a) Average segment density profiles of tangentially
jointed Lennard-Jones 8-mers before
(dashed line, $\rho^*_b=2.0244243\times 10^{-07}$)
and after (solid line, $\rho^*_b=2.0244424\times 10^{-07}$)
the prewetting transition
evaluated for $T^*=1.1$. The dashed profile was multiplied by 500. 
(b) Average segment density profiles of tangentially
jointed Lennard-Jones 8-mers before (dashed line, $\rho^*_b=0.0015$)
and after (solid lines, from left to right for
$\rho^*_b=0.0016, 0.0017, 0.0018, 0.0019, 0.0020, 0.0021, 0.00215$,
and $0.002156$, respectively) the prewetting transition
evaluated for $T^*=1.85$. Dotted line in both figures marks
bulk segment density of the liquid phase.

\textbf{Fig. 4.} Segment adsorption isotherms of tangentially
jointed Lennard-Jones 8-mers adsorbed on a solid planar surface
with $\varepsilon_{gs}^*=6$ (circles) plotted as a function
of the logarithm of the deviation from bulk coexistence 
$\log(\rho^*_b-\rho^*_{b,c})$
evaluated at two reduced temperatures: $T^*=1.1$ (part a), and $T^*=1.85$
(part b). Solid line in both parts has a slope -1/3.

\textbf{Fig. 5.} (a) Middle (solid lines)
and end (dashed lines) segment density profiles of tangentially
jointed Lennard-Jones 8-mers before
($\rho^*_b=2.0244243\times 10^{-07}$)
and after ($\rho^*_b=2.0244424\times 10^{-07}$)
the prewetting transition evaluated for $T^*=1.1$.
The profiles before the prewetting transition were multiplied by 500.
(b) Middle (solid lines) and end (dashed lines)
segment density profiles of tangentially
jointed Lennard-Jones 8-mers  
evaluated for $T^*=1.85$ and for $\rho^*_b=0.00215$.

\textbf{Fig. 6.} The surface phase diagrams of tangentially
jointed Lennard-Jones dimers (a) and 8-mers (b) plotted in the
reduced bulk density -- reduced temperature representation.
Shown are the prewetting lines (solid lines), the gas branch of
the corresponding bulk phase diagrams (dashed lines), the bulk critical
points (filled circles), the surface critical points (open circles)
and the wetting points (open triangles). The corresponding
temperatures are marked in the Figure.

\textbf{Fig. 7} The dependence of the reduced bulk critical
temperature $T^*_c$ (filled circles), the rescaled surface
critical temperature $T^*_{sc}/T^*_c$ (open circles), and
the rescaled wetting temperature $T^*_{w}/T^*_c$
(open triangles) on the chain length $M$. Dotted and 
dashed lines serve as a guide to the eye. Arrows point
to the relevant temperature axis.

\textbf{Fig. 8} The dependence of the reduced surface
critical temperature $T^*_{sc}$ (open circles), and the
reduced wetting temperature $T^*_{w}$ (open triangles)
on the reduced segment-surface energy parameter $\varepsilon_{gs}^*$
evaluated for the constant chain length $M=4$.

\newpage
\begin{figure}[tbp]
\includegraphics[width=16cm,clip]{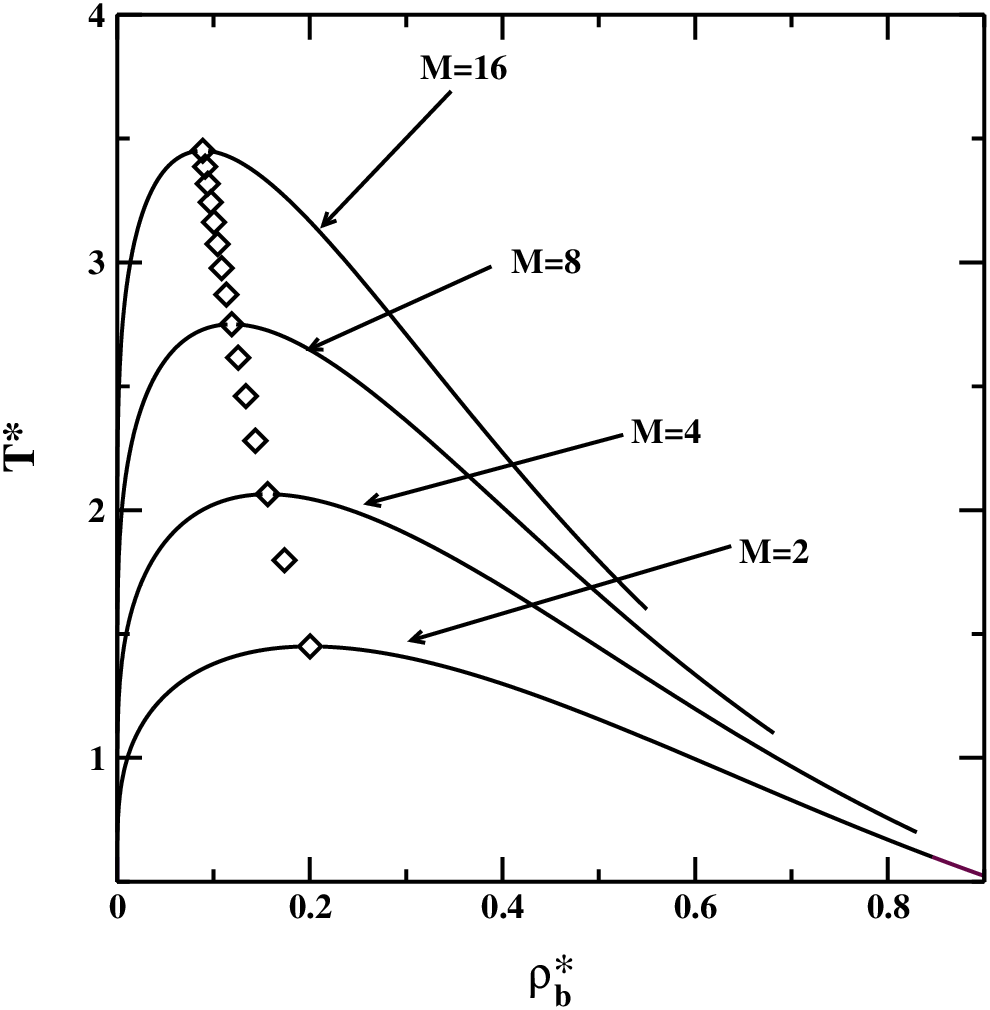}
\caption{}
\label{fig:1}
\end{figure}

\begin{figure}[tbp]
\includegraphics[width=16cm,clip]{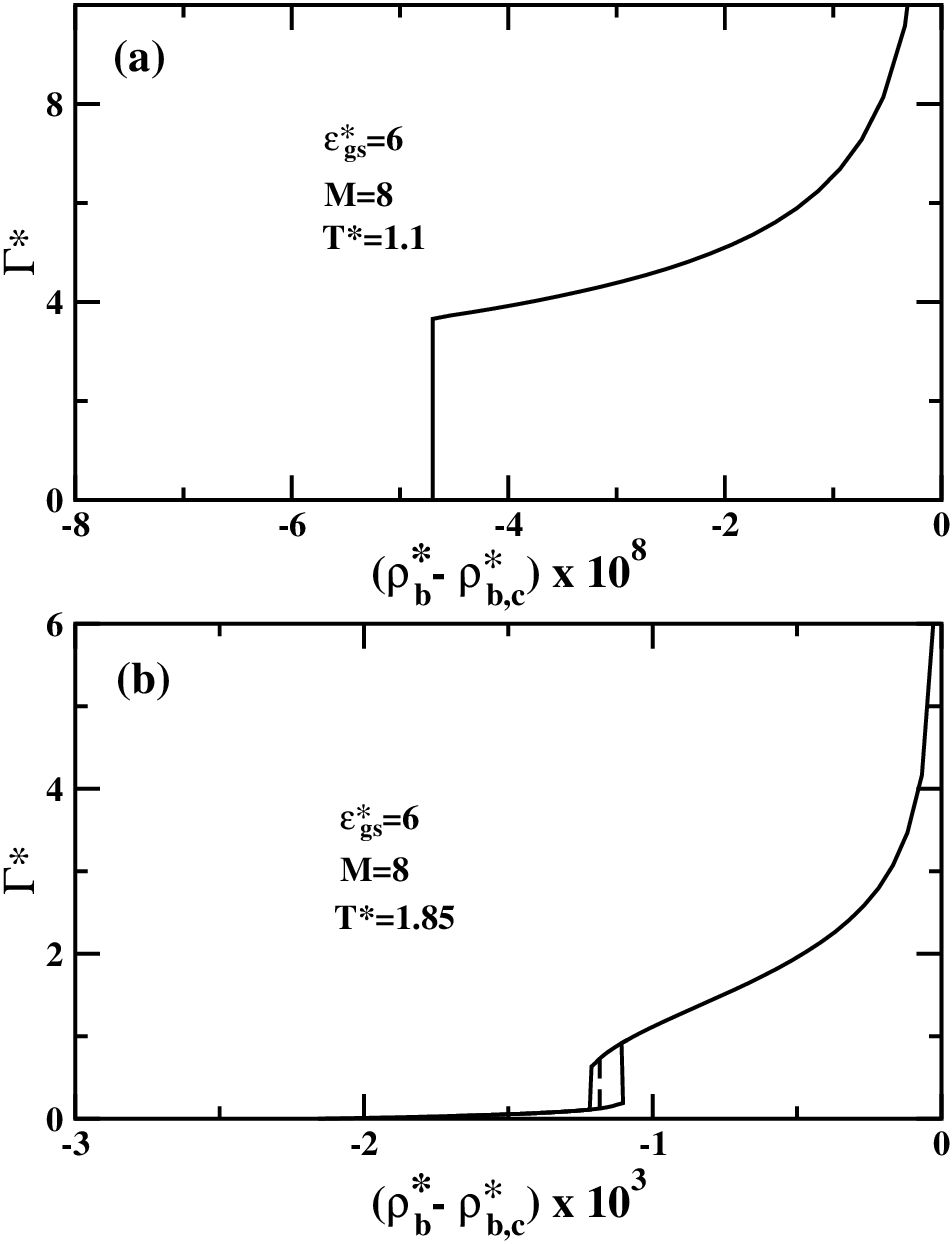}
\caption{}
\label{fig:2}
\end{figure}

\begin{figure}[tbp]
\includegraphics[width=16cm,clip]{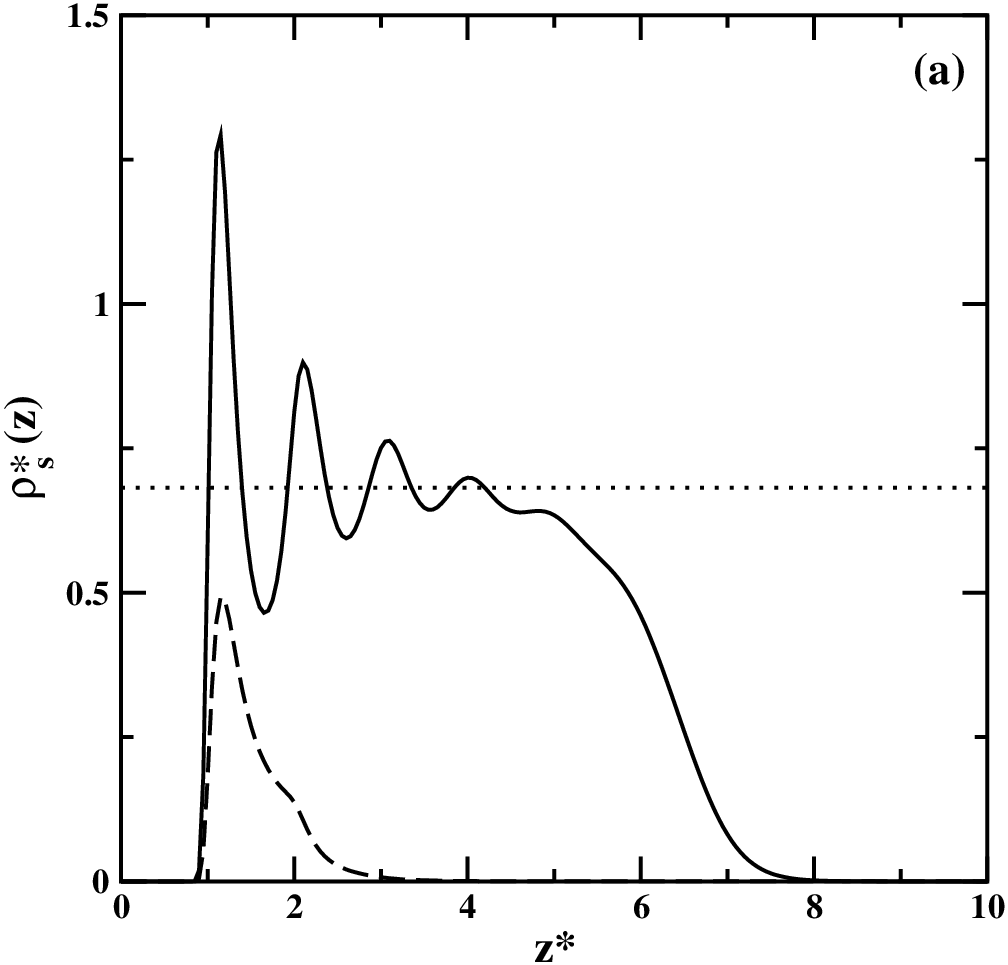}
\caption{}
\label{fig:3}
\end{figure}

\begin{figure}[tbp]
\includegraphics[width=16cm,clip]{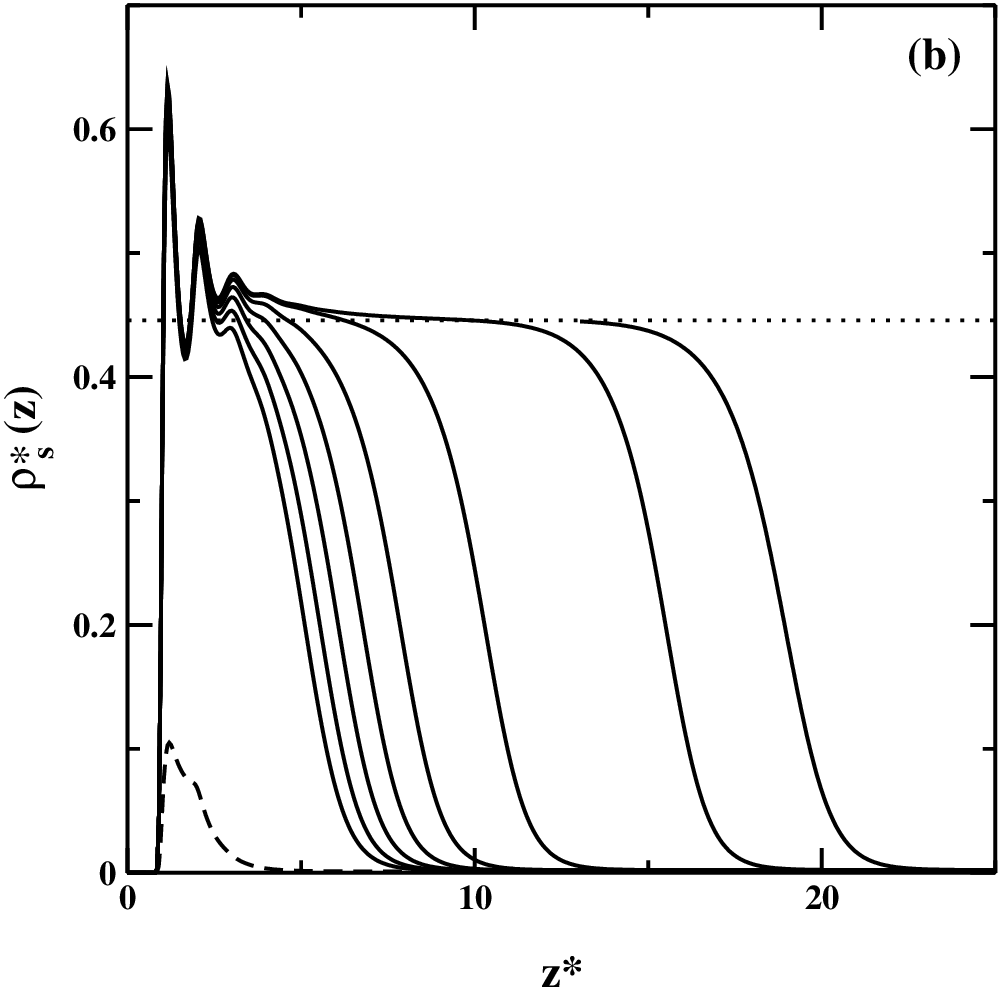}
\end{figure}

\begin{figure}[tbp]
\includegraphics[width=16cm,clip]{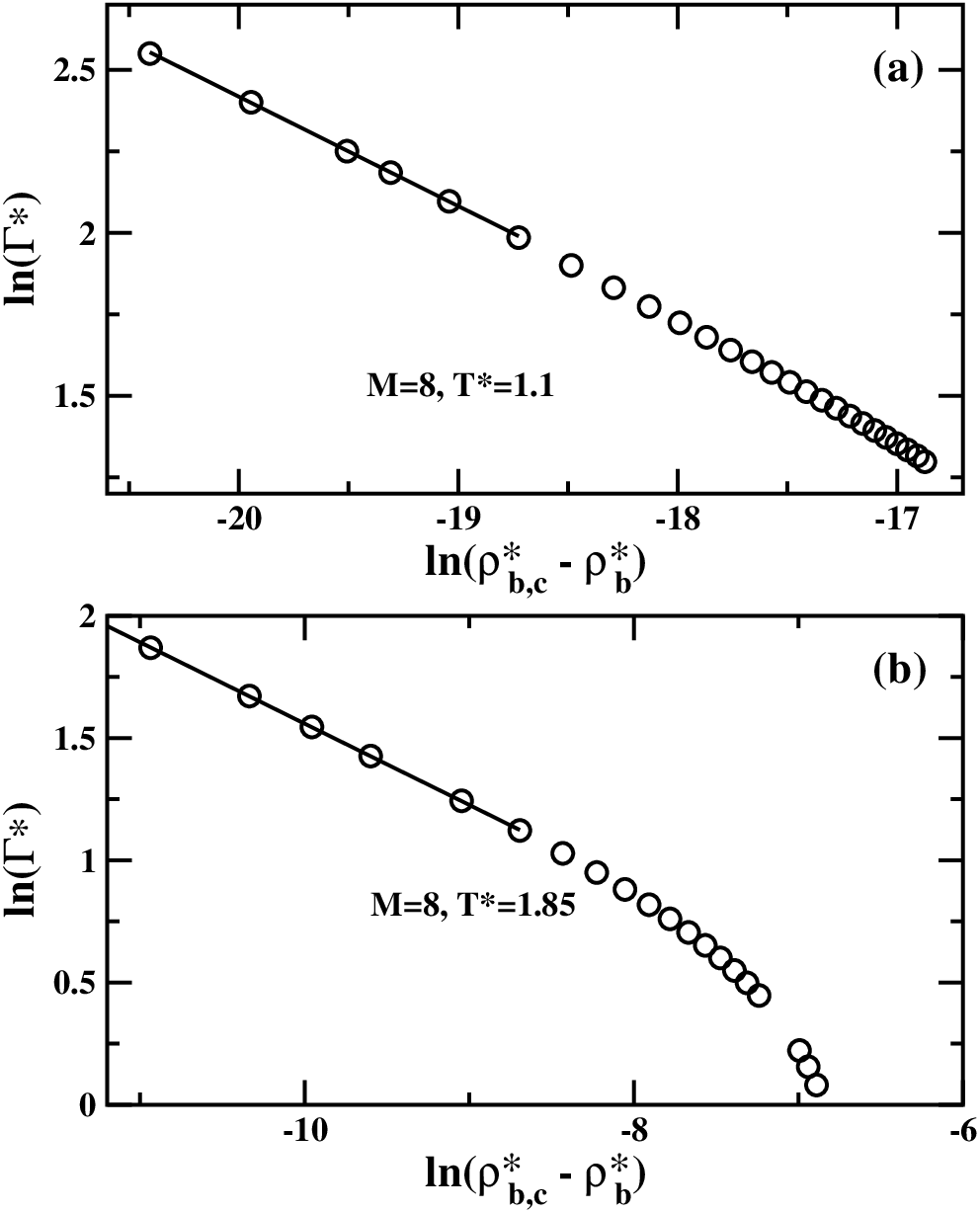}
\caption{}
\label{fig:4}
\end{figure}

\begin{figure}[tbp]
\includegraphics[width=16cm,clip]{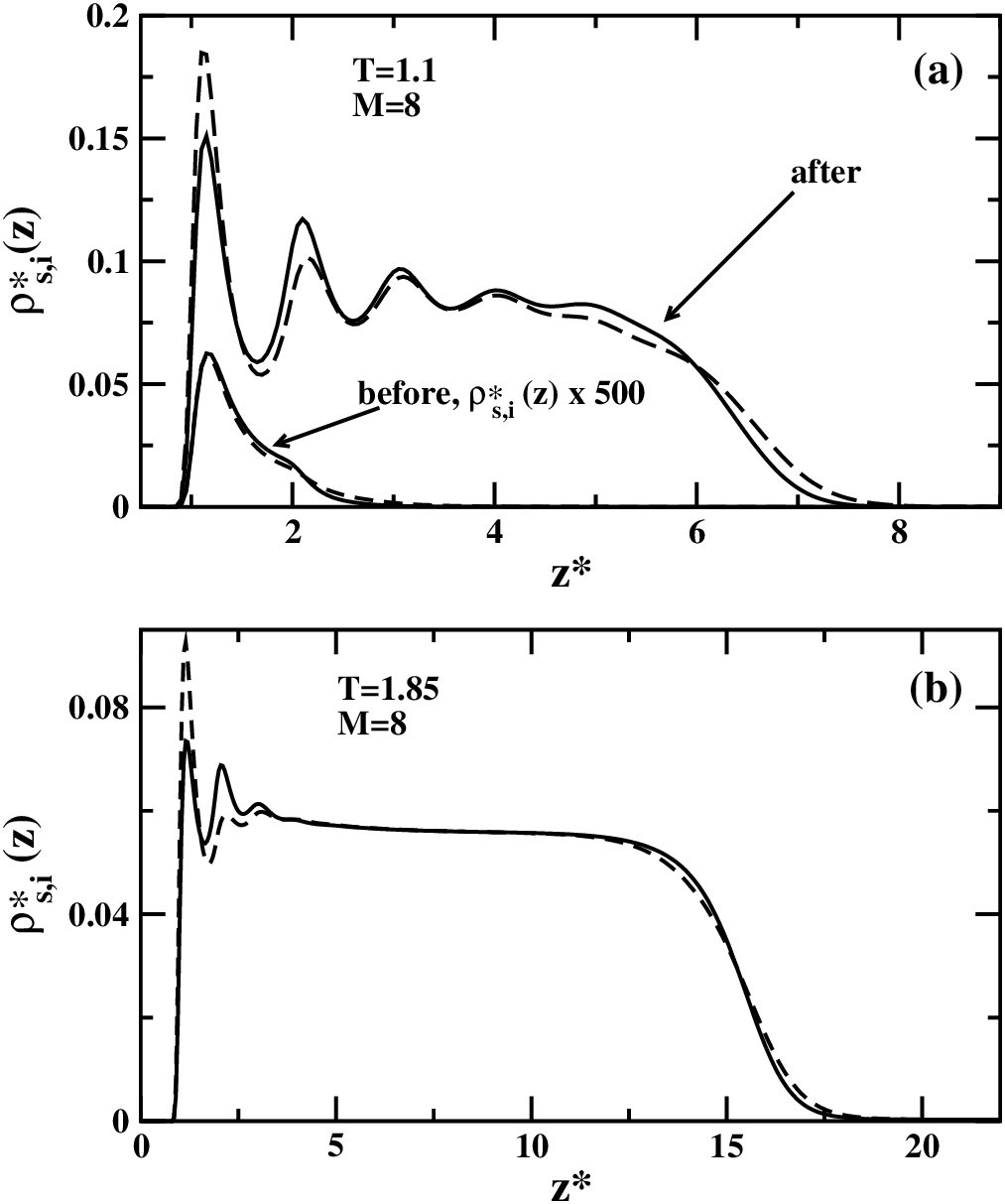}
\caption{}
\label{fig:5}
\end{figure}

\begin{figure}[tbp]
\includegraphics[width=16cm,clip]{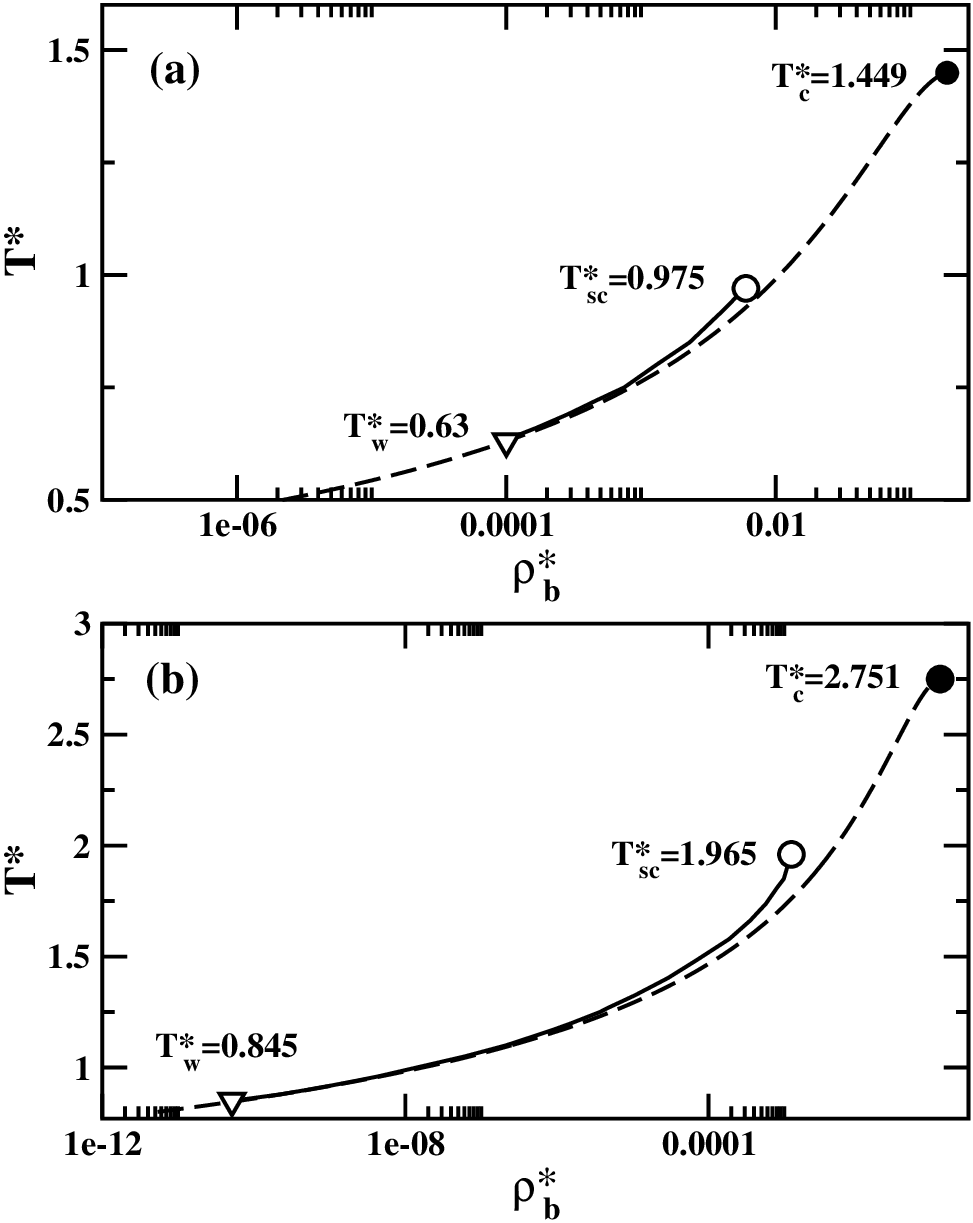}
\caption{}
\label{fig:6}
\end{figure}

\begin{figure}[tbp]
\includegraphics[width=16cm,clip]{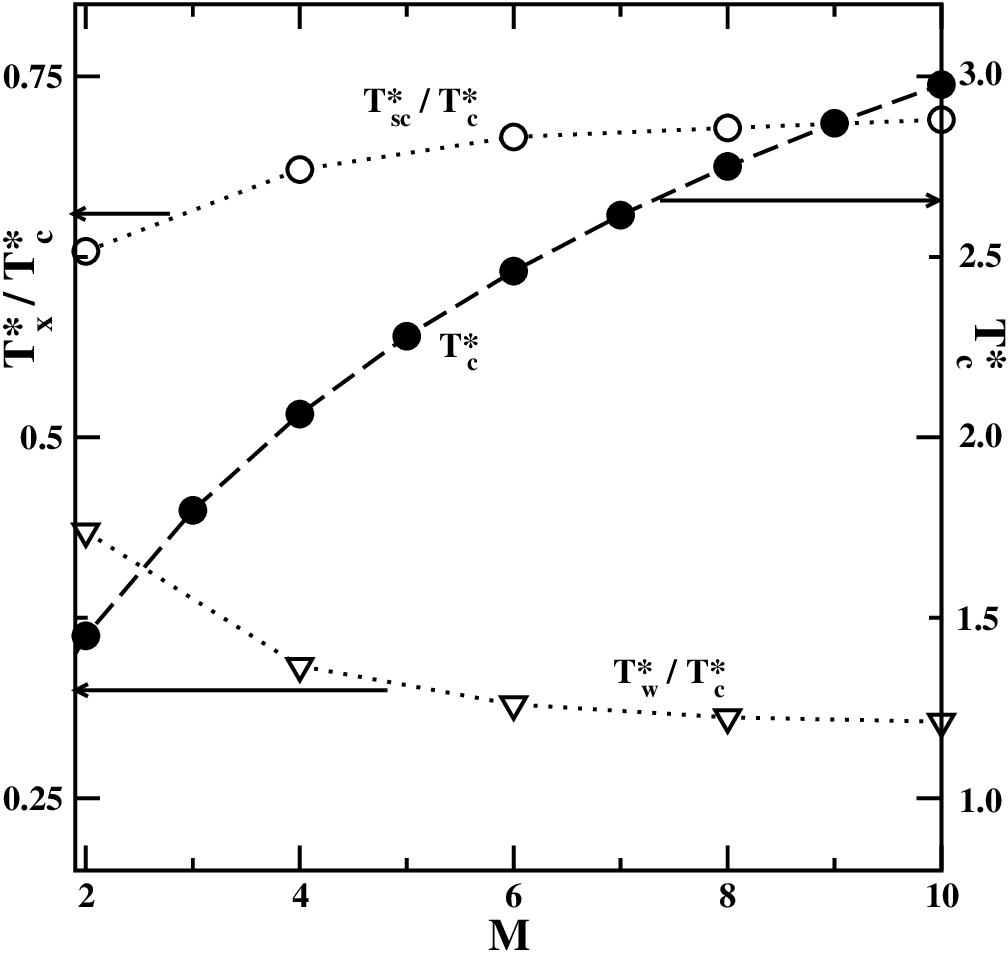}
\caption{}
\label{fig:7}
\end{figure}

\begin{figure}[tbp]
\includegraphics[width=16cm,clip]{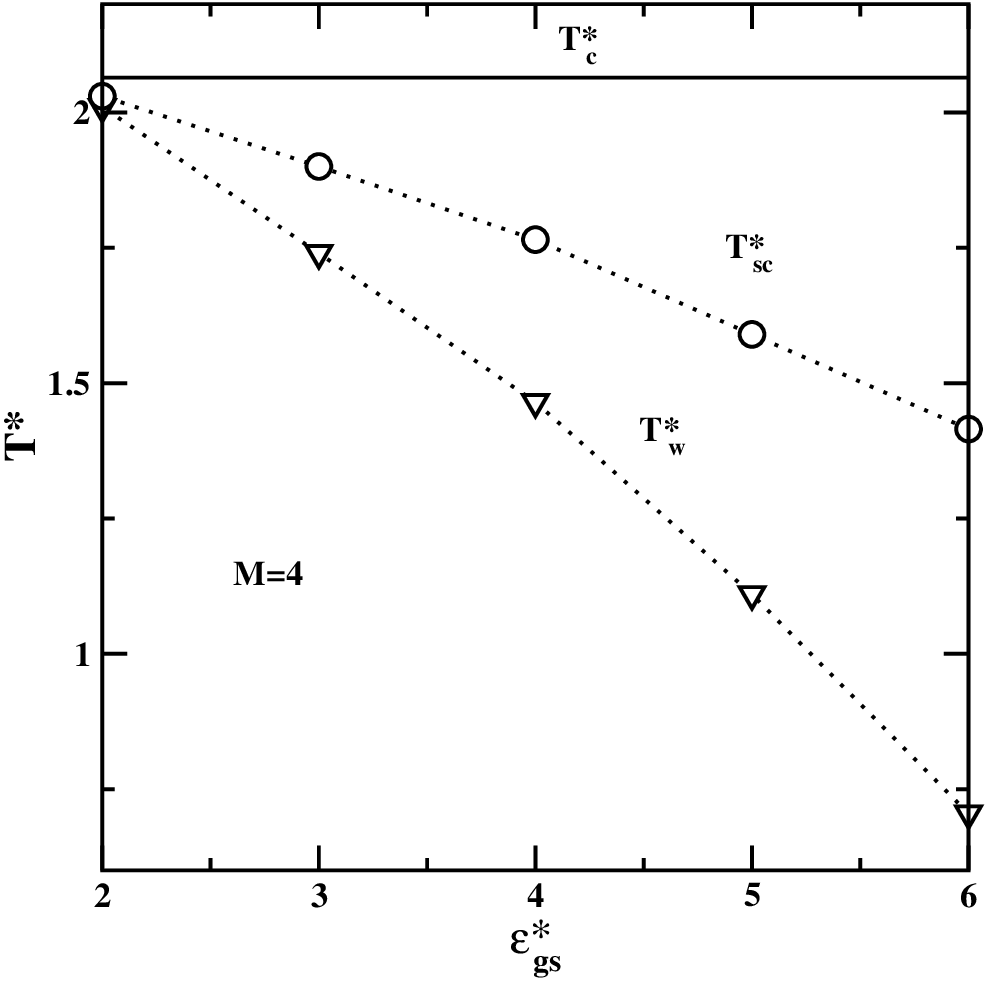}
\caption{}
\label{fig:8}
\end{figure}
\end{document}